\pdfoutput=1
\documentclass[11pt]{article}
\usepackage[T1]{fontenc}
\usepackage[utf8]{inputenc}
\usepackage{lmodern}
\usepackage[margin=0.82in]{geometry}
\usepackage{amsmath,amssymb,bm}
\usepackage{graphicx}
\usepackage{booktabs,tabularx,array}
\usepackage{microtype}
\usepackage{xurl}
\usepackage[hidelinks]{hyperref}
\usepackage{placeins}
\usepackage{enumitem}
\newcolumntype{Y}{>{\raggedright\arraybackslash}X}
\newcolumntype{C}{>{\centering\arraybackslash}X}
\newcolumntype{R}{>{\raggedleft\arraybackslash}X}
\title{Quantum-Classical Effective Fragment Potential Embedding for Condensed-Phase Quantum Chemistry}
\author{Federico Zahariev$^{1,2}$, Vassiliki-Alexandra Glezakou$^{3}$, and Mark S. Gordon$^{1,2}$}
\date{}
\begin{document}
\maketitle
\begin{center}
\small $^{1}$Department of Chemistry, Iowa State University, Ames, Iowa 50011, USA\\$^{2}$Ames National Laboratory, Ames, Iowa 50011, USA\\$^{3}$Chemical Sciences Division, Oak Ridge National Laboratory, Oak Ridge, Tennessee 37831, USA
\end{center}
\vspace{0.5em}
\begin{abstract}
Quantum simulation of chemistry in realistic environments is constrained by the orbital cost of explicit solvent, ions, and other surroundings. We present the quantum-classical effective fragment potential (Q-EFP) method, in which a chemically active region is treated with a quantum algorithm and the environment is represented by EFP in GAMESS. Coulomb and polarization potentials enter the active-region one-electron Hamiltonian, while EFP--EFP and short-range environment contributions are assembled classically. Statevector UCCSD/STO-3G benchmarks for LiH in a mixed water--methanol environment, H$_2$O in a five-water environment, and BeH$_2$ in an ammonium--nitrate environment differ from matched classical CCSD/EFP calculations by 0.03, 0.38, and 0.01 kcal mol$^{-1}$, respectively. The active calculations require 4--8 qubits, compared with estimated full-system counts of approximately 84--246 qubits, corresponding to register reductions of about 10--35-fold. These proof-of-concept results validate the embedded Hamiltonian and Q-GAMESS workflow while separating environmental size from quantum-register size. Q-EFP is complementary to real-space Q-EFMO fragmentation and virtual-orbital Q-FVO reduction, enabling a layered route to larger solvated systems.
\end{abstract}
\section{Introduction}
Quantum algorithms provide a natural representation of correlated electronic wave functions, but chemically realistic calculations remain constrained by the size and depth of the quantum subproblem. Variational algorithms such as the variational quantum eigensolver (VQE) reduce coherent-depth requirements by combining quantum expectation values with classical optimization, yet the active orbital spaces accessible to present hardware and high-fidelity simulators remain modest \cite{preskill2018,mcardle2020,cao2019,bauer2020,peruzzo2014,mcclean2016,tilly2022}. This limitation is especially acute for condensed-phase chemistry, where explicit solvent, counterions, surfaces, or protein environments can contribute many more orbitals than the chemically active solute.

Embedding methods address this imbalance by assigning the strongly correlated or chemically changing region to a high-level solver and treating the environment with a lower-cost model. Classical fragmentation and quantum-mechanics/molecular-mechanics strategies have demonstrated that much of the long-range environmental response can be represented without placing every environmental orbital in the correlated wave function \cite{gordon2012}. For quantum computing, this separation has a direct resource consequence: if the environment enters as a potential rather than as additional second-quantized orbitals, its size need not determine the quantum-register size.

The Effective Fragment Potential (EFP) method is particularly attractive for this purpose. EFP represents intermolecular interactions through distributed electrostatics, many-body polarization, dispersion, exchange-repulsion, and charge-transfer contributions derived from fragment calculations and overlap-based expressions \cite{jensen1996,gordon2001,gordon2007}. The general EFP2 model is applicable to arbitrary molecular fragments, whereas the historically earlier EFP1 model is water-specific and uses a distinct parameterization. EFP has been combined with ground- and excited-state electronic-structure methods and has reproduced solvent-induced spectral shifts at a small fraction of the cost of fully explicit quantum calculations \cite{yoo2008,sok2011}.

Here we formulate and test a quantum-classical EFP embedding workflow, denoted Q-EFP. The active quantum-mechanical (QM) region is solved with VQE/UCCSD, the environment is evaluated by EFP in GAMESS, and the two contributions are assembled through Q-GAMESS. The present study has three aims: to define the embedded Hamiltonian and energy bookkeeping; to demonstrate an automated GAMESS-to-quantum-backend workflow; and to quantify accuracy and register-size reductions for three small statevector benchmarks. The results are proof-of-concept tests of the embedding and software path, not benchmarks of noisy hardware or converged basis-set thermochemistry.

Q-EFP occupies one layer of a broader decomposition hierarchy. Q-EFP removes a large environment from the quantum register by replacing it with a first-principles embedding potential. The companion Q-EFMO workflow partitions an extended molecular cluster into real-space monomers and near-field dimers, and virtual-orbital fragmentation (Q-FVO) can further reduce the active virtual space within any quantum subproblem. These approaches are therefore complementary rather than competing: environment embedding, real-space fragmentation, and orbital-space fragmentation act on different sources of quantum-resource growth.

\section{Method}
\subsection{Q-EFP partition}
\begin{figure}[tb]
\centering
\includegraphics[width=0.98\textwidth]{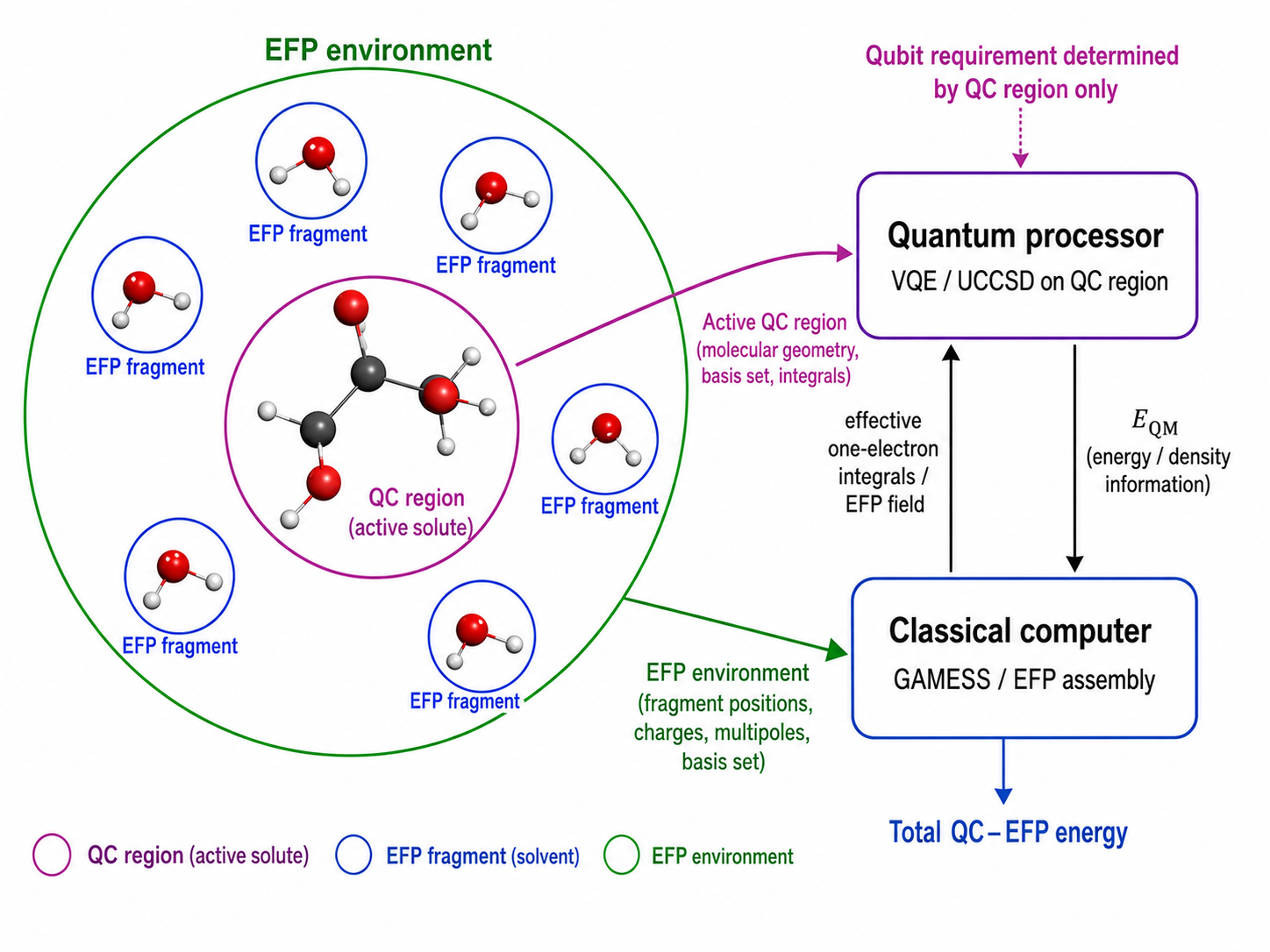}
\caption{Q-EFP embedding and information flow. Only the chemically active region is mapped to qubits and treated with VQE/UCCSD. The EFP environment is handled in GAMESS; its electrostatic and polarization potentials modify the active-region one-electron integrals, while the quantum energy or density information returns to the classical assembly. Consequently, the qubit requirement is determined by the active region rather than by the number of EFP fragments.}
\label{fig:qc_embedding}
\end{figure}

Q-EFP partitions the molecular system into one active QM region and a set of EFP fragments. Figure~\ref{fig:qc_embedding} summarizes the information flow. The environment produces electrostatic and induced-polarization potentials that modify the one-electron operator of the QM region. The resulting embedded Hamiltonian is mapped to qubits and solved variationally. EFP--EFP interactions and the environment terms not represented in the embedded one-electron operator are retained in the classical energy assembly. Because the fragment orbitals do not enter the second-quantized QM Hamiltonian, adding EFP fragments changes the classical workload but does not add qubits.

\subsection{Effective fragment potential environment}
\begin{figure}[tb]
\centering
\includegraphics[width=0.92\textwidth]{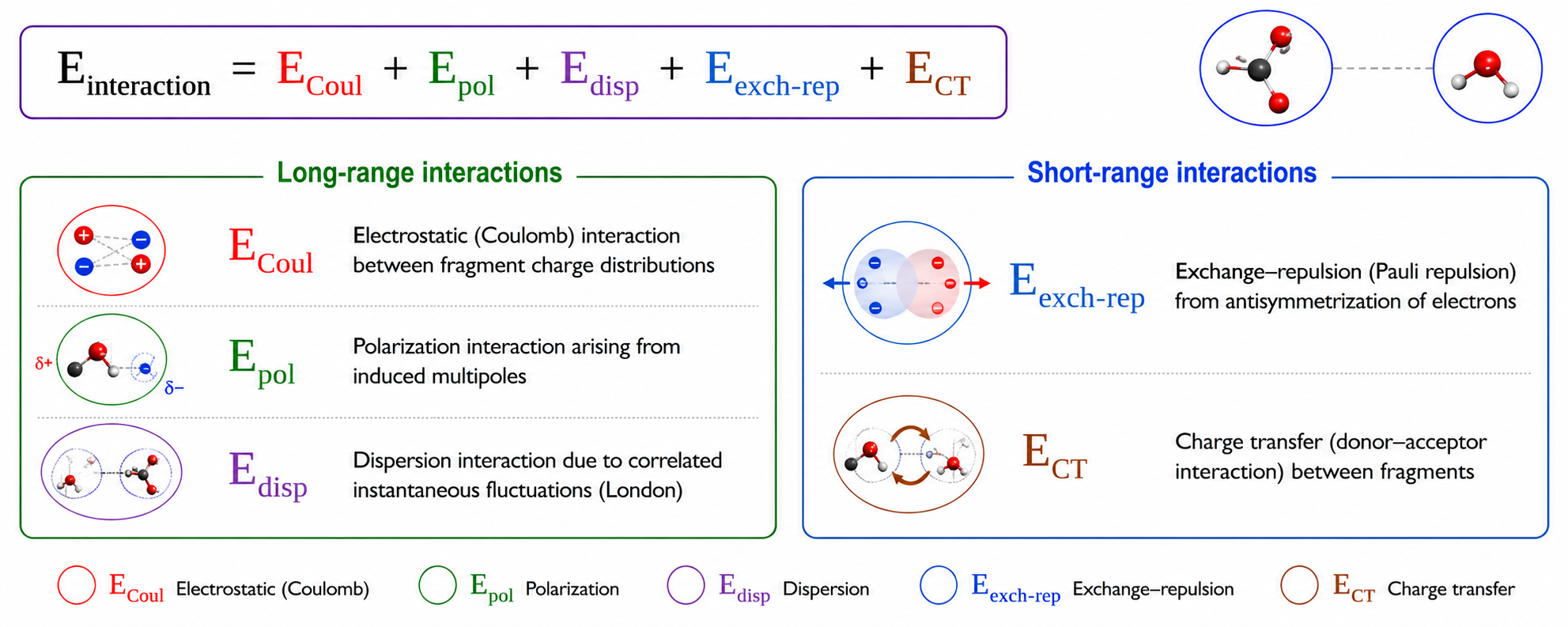}
\caption{Physical decomposition of the EFP interaction energy. Coulomb, polarization, and dispersion provide long-range contributions, whereas exchange-repulsion and charge transfer are short-range and overlap-dependent. The diagram summarizes the general EFP2 interpretation; the water-specific EFP1 model uses its established historical parameterization.}
\label{fig:efp_components}
\end{figure}

In the general EFP2 formulation, the interaction energy is separated into physically interpretable terms,

\begin{equation}
E_{\mathrm{int}}^{\mathrm{EFP}}=E_{\mathrm{elst}}+E_{\mathrm{pol}}+E_{\mathrm{disp}}+E_{\mathrm{exch-rep}}+E_{\mathrm{CT}}.
\label{eq:1}
\end{equation}
Distributed multipoles through octupole order describe anisotropic electrostatics. Distributed polarizability tensors are iterated to self-consistency, which introduces a many-body induction response across the fragments and the QM density. Dispersion is represented by damped inverse-power terms, whereas exchange-repulsion and charge transfer are short-range, overlap-dependent contributions. The water-specific EFP1 benchmark below uses the established EFP1 parameterization rather than assuming that every EFP1 term has the same derivation as EFP2. Figure~\ref{fig:efp_components} emphasizes this distinction.

\subsection{Embedded Hamiltonian and quantum solver}
Let $h_{pq}^{(0)}$ denote the one-electron integrals of the isolated QM region in an orthonormal molecular-orbital basis. The EFP electrostatic and polarization potentials modify these integrals according to

\begin{equation}
h_{pq}^{\mathrm{eff}}=h_{pq}^{(0)}+\langle p|v_{\mathrm{elst}}^{\mathrm{EFP}}+v_{\mathrm{pol}}^{\mathrm{EFP}}|q\rangle.
\label{eq:2}
\end{equation}
The embedded electronic Hamiltonian is then

\begin{equation}
\hat H_{\mathrm{emb}}=\sum_{pq}h_{pq}^{\mathrm{eff}}a_p^\dagger a_q+\frac{1}{2}\sum_{pqrs}g_{pqrs}a_p^\dagger a_q^\dagger a_s a_r.
\label{eq:3}
\end{equation}
GAMESS supplies the modified integrals and the classical EFP energy components. The total energy can be written compactly as

\begin{equation}
E_{\mathrm{Q\mbox{-}EFP}}=\langle\Psi(\bm\theta)|\hat H_{\mathrm{emb}}|\Psi(\bm\theta)\rangle+E_{\mathrm{EFP}}^{\mathrm{classical,corr}},
\label{eq:4}
\end{equation}
where $E_{\mathrm{EFP}}^{\mathrm{classical,corr}}$ denotes the EFP--EFP energy, nuclear--fragment terms, short-range corrections, and the polarization bookkeeping required to avoid double counting. In the implementation, this correction is taken from GAMESS rather than reconstructed independently.

\begin{figure}[tb]
\centering
\includegraphics[width=0.88\textwidth]{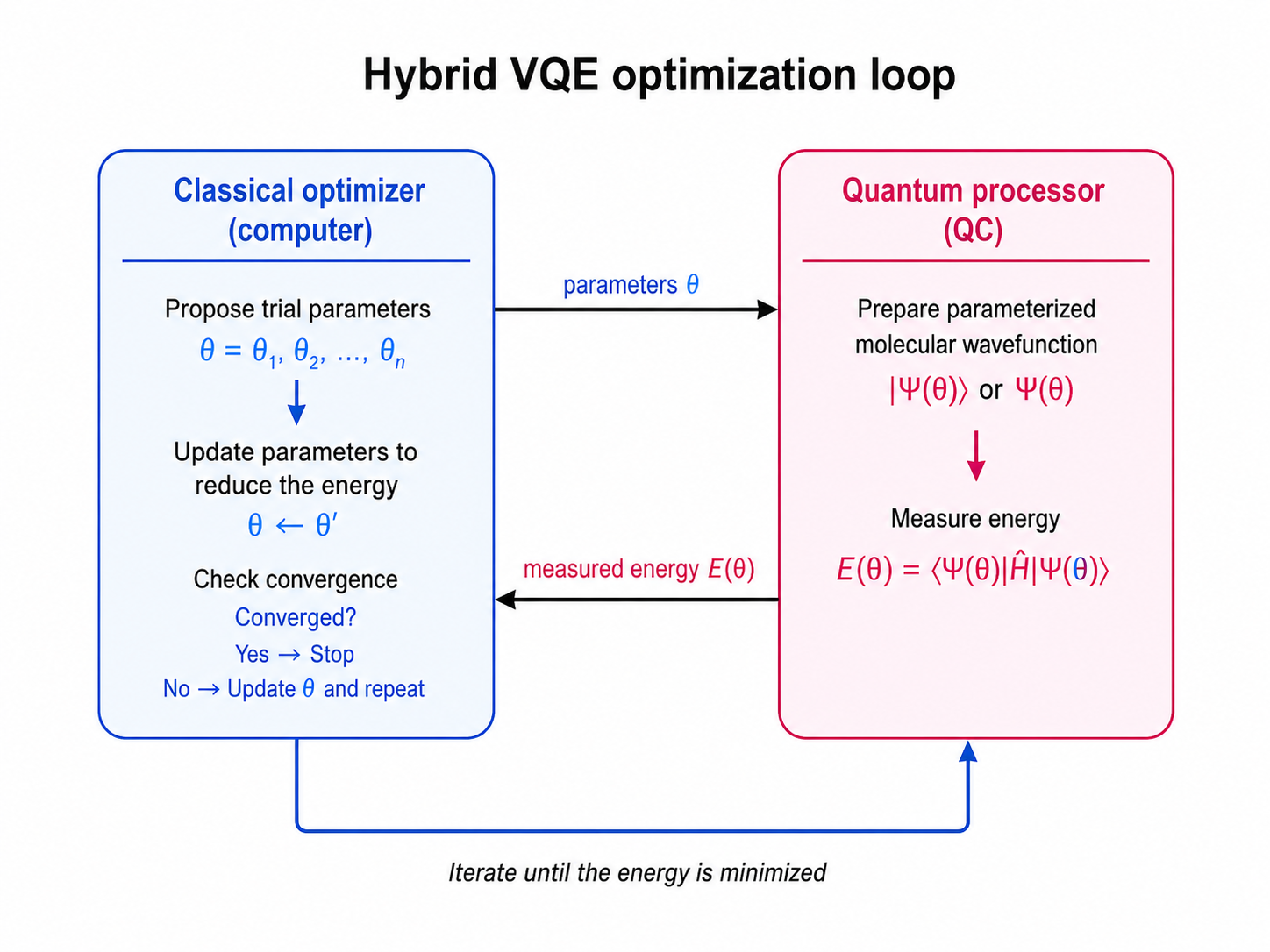}
\caption{VQE feedback loop. A classical optimizer proposes the UCCSD parameters, the quantum backend prepares the trial state and evaluates the embedded Hamiltonian, and the resulting energy (and, when available, gradients) is returned to the optimizer.}
\label{fig:vqe_loop}
\end{figure}

The VQE trial state uses the unitary coupled-cluster singles-and-doubles (UCCSD) ansatz \cite{romero2018,anand2022},

\begin{equation}
|\Psi(\bm\theta)\rangle=\exp(\hat T-\hat T^\dagger)|\Phi_0\rangle,\qquad \hat T=\sum_{ia}t_i^a a_a^\dagger a_i+\frac{1}{4}\sum_{ijab}t_{ij}^{ab}a_a^\dagger a_b^\dagger a_j a_i.
\label{eq:5}
\end{equation}
Here $i,j$ and $a,b$ label occupied and virtual spin orbitals, respectively. A classical optimizer updates the amplitudes from measured or simulated energy expectation values until convergence (Figure~\ref{fig:vqe_loop}). The present calculations use statevector simulation, so they isolate the electronic-structure and workflow consistency from shot noise and device errors.

\subsection{Quantum-resource decoupling}
Before symmetry tapering or other encoding-specific reductions, the register size is

\begin{equation}
N_q=2M_{\mathrm{active}},
\label{eq:6}
\end{equation}
where $M_{\mathrm{active}}$ is the number of active spatial orbitals after frozen-core and virtual-space reductions. The number of EFP fragments is absent from Eq.~\ref{eq:6}. It can affect the cost of self-consistent polarization and classical pair interactions, but not the number of qubits for a fixed QM active space. Jordan--Wigner and Bravyi--Kitaev mappings were available in Q-GAMESS \cite{jordan1928,bravyi2002}.

\section{Implementation and computational details}
The Q-EFP workflow is implemented through Q-GAMESS, which connects GAMESS electronic-structure and EFP calculations \cite{barca2020} with quantum-computing software. The calculation proceeds in four stages. First, GAMESS generates the QM-region geometry, basis information, one- and two-electron integrals, and the EFP parameters or fragment data for the environment. Second, Coulomb and self-consistent polarization contributions are folded into the one-electron integrals before the qubit Hamiltonian is constructed. Third, the embedded active-space Hamiltonian is mapped to qubits and solved with statevector VQE/UCCSD. Fourth, the quantum expectation value is combined with nuclear, EFP--EFP, dispersion, exchange-repulsion, charge-transfer, and polarization-bookkeeping terms from GAMESS to obtain the total Q-EFP energy. The modular interface can target Qiskit, PennyLane, and OpenFermion- or PySCF-compatible workflows; the precise package versions and Q-GAMESS commit are recorded with the corresponding input files \cite{qgamess2025,openfermion2020,pyscf2020,qiskit2024}.

Three systems were selected to exercise distinct environments: LiH embedded in three water and three methanol fragments using EFP2, one water molecule embedded in five EFP1 water fragments, and BeH$_2$ embedded in four ammonium and four nitrate EFP2 fragments. All calculations used the STO-3G basis \cite{hehre1969} and statevector simulation. One chemically inert core spatial orbital was frozen in each active region. For LiH, two additional virtual orbitals were removed from the active space, leaving 4 qubits; the water and BeH$_2$ calculations used 8 qubits. The matched reference for each benchmark is a classical CCSD/EFP calculation \cite{purvis1982} with the same basis, geometry, frozen-core convention, and EFP environment. Thus the reported differences isolate the current embedding and solver pathway rather than basis-set, geometry, or device-noise errors.

The VQE optimizer, convergence criteria, orbital indices, EFP iteration settings, and package versions are documented in the corresponding input files. The present calculations were noise-free statevector tests and did not include finite-shot sampling, device noise, or error mitigation.

\section{Results}
\subsection{LiH in a mixed water--methanol environment}
Table~\ref{tab:lih} compares isolated and embedded LiH. The isolated UCCSD and CCSD values agree to the reported precision. In the EFP2 environment, the absolute difference is $0.03$ kcal mol$^{-1}$. The fixed-geometry environment-induced energy lowering is $41.99$ kcal mol$^{-1}$ for CCSD/EFP2 and $42.02$ kcal mol$^{-1}$ for Q-EFP2.

\begin{table}[tb]
\centering
\small
\caption{Energies for LiH with and without the EFP2 environment (3 H$_2$O + 3 CH$_3$OH).}
\label{tab:lih}
\begin{tabularx}{\textwidth}{@{}YCC@{}}
\toprule
Method & LiH, isolated (hartree) & LiH + EFP2 (hartree) \\
\midrule
GAMESS CCSD & $-7.843260$ & $-7.910175$ \\
QC UCCSD & $-7.843260$ & $-7.910223$ \\
Absolute difference (kcal mol$^{-1}$) & $0.00$ & $0.03$ \\
\bottomrule
\end{tabularx}
\vspace{0.35em}\parbox{0.98\textwidth}{\footnotesize STO-3G; Li $1s$ core frozen; two high-energy virtual orbitals removed; 4-qubit active space.}
\end{table}
\subsection{Water in a hydrogen-bonded cluster}
For one water molecule embedded in five EFP1 water fragments, the isolated calculations differ by $0.02$ kcal mol$^{-1}$ and the embedded calculations by $0.38$ kcal mol$^{-1}$ (Table~\ref{tab:water}). The corresponding energy lowerings are $36.58$ and $36.98$ kcal mol$^{-1}$ for CCSD/EFP1 and Q-EFP1, respectively. The larger difference than in the other two tests remains below $0.5$ kcal mol$^{-1}$, but the present data do not isolate whether it originates from polarization convergence, orbital reduction, or another implementation detail.

\begin{table}[tb]
\centering
\small
\caption{Energies for H$_2$O with and without five EFP1 water fragments.}
\label{tab:water}
\begin{tabularx}{\textwidth}{@{}YCC@{}}
\toprule
Method & H$_2$O, isolated (hartree) & H$_2$O + EFP1 (hartree) \\
\midrule
GAMESS CCSD & $-74.966645$ & $-75.024939$ \\
QC UCCSD & $-74.966613$ & $-75.025545$ \\
Absolute difference (kcal mol$^{-1}$) & $0.02$ & $0.38$ \\
\bottomrule
\end{tabularx}
\vspace{0.35em}\parbox{0.98\textwidth}{\footnotesize STO-3G; O $1s$ core frozen; final active space 8 qubits.}
\end{table}
\subsection{BeH$_2$ in an ionic environment}
The charged-fragment workflow was tested with BeH$_2$ embedded in four NH$_4^+$ and four NO$_3^-$ EFP2 fragments. The isolated and embedded UCCSD values both differ from the corresponding CCSD values by $0.01$ kcal mol$^{-1}$ (Table~\ref{tab:beh2}). The environment-induced energy lowering is approximately $430.27$ kcal mol$^{-1}$ at the fixed geometry. This large shift is a property of the chosen ionic arrangement and should not be interpreted as a transferable solvation energy.

\begin{table}[tb]
\centering
\small
\caption{Energies for BeH$_2$ with and without the EFP2 ionic environment (4 NH$_4^+$ + 4 NO$_3^-$).}
\label{tab:beh2}
\begin{tabularx}{\textwidth}{@{}YCC@{}}
\toprule
Method & BeH$_2$, isolated (hartree) & BeH$_2$ + EFP2 (hartree) \\
\midrule
GAMESS CCSD & $-15.569741$ & $-16.255420$ \\
QC UCCSD & $-15.569757$ & $-16.255436$ \\
Absolute difference (kcal mol$^{-1}$) & $0.01$ & $0.01$ \\
\bottomrule
\end{tabularx}
\vspace{0.35em}\parbox{0.98\textwidth}{\footnotesize STO-3G; Be core frozen; final active space 8 qubits.}
\end{table}
\subsection{Accuracy and register-size summary}
Across the three embedded calculations, the largest absolute difference from the matched CCSD/EFP reference is $0.38$ kcal mol$^{-1}$ (Table~\ref{tab:accuracy}). Relative to the magnitude of the environment-induced energy shift, the deviations range from approximately $0.002\%$ to $1.03\%$. These percentages are descriptive ratios, not formal statistical accuracies.

\begin{table}[tb]
\centering
\small
\caption{Summary of Q-EFP differences from matched classical CCSD/EFP calculations.}
\label{tab:accuracy}
\begin{tabularx}{\textwidth}{@{}YCCC@{}}
\toprule
System & Absolute difference (kcal mol$^{-1}$) & $|\Delta E_{\mathrm{env}}|$ (kcal mol$^{-1}$) & Relative deviation (\%) \\
\midrule
LiH + mixed solvent & $0.03$ & $42.0$ & $0.071$ \\
H$_2$O + water cluster & $0.38$ & $37.0$ & $1.03$ \\
BeH$_2$ + ionic environment & $0.01$ & $430.3$ & $0.0023$ \\
\bottomrule
\end{tabularx}
\end{table}
Table~\ref{tab:qubits} compares the active Q-EFP registers with full-system STO-3G spin-orbital counts. The environment contains 15--36 atoms, yet the quantum calculations require only 4--8 qubits. The resulting register reductions are approximately 10--35-fold for the examples considered. Larger solvent shells would increase the full-system count while leaving the Q-EFP register unchanged, provided that the QM active region is held fixed.

\begin{table}[tb]
\centering
\small
\caption{Estimated qubit requirements for Q-EFP and an explicit STO-3G treatment of all atoms.}
\label{tab:qubits}
\begin{tabularx}{\textwidth}{@{}YCCCC@{}}
\toprule
System & Q-EFP qubits & Estimated full-system qubits & Reduction & Environment atoms \\
\midrule
LiH + 6 solvent molecules & $4$ & $\sim138$ & $\sim35\times$ & $27$ \\
H$_2$O + 5 H$_2$O & $8$ & $\sim84$ & $\sim10\times$ & $15$ \\
BeH$_2$ + 8 ions & $8$ & $\sim246$ & $\sim31\times$ & $36$ \\
\bottomrule
\end{tabularx}
\vspace{0.35em}\parbox{0.98\textwidth}{\footnotesize Full-system estimates use one qubit per spin orbital and no frozen-core approximation.}
\end{table}
\section{Discussion}
The numerical agreement demonstrates internal consistency between the embedded UCCSD pathway and the matched classical CCSD/EFP calculations for the selected active spaces. It does not establish basis-set convergence, broad transferability, or noisy-hardware performance. The STO-3G basis captures only a limited portion of dynamical correlation, and statevector simulation omits finite-shot uncertainty, gate error, and error-mitigation bias.

The principal advantage of Q-EFP is therefore architectural rather than tied to the absolute values in these small tests: environmental degrees of freedom are represented by potentials and classical interaction terms instead of qubit orbitals. The approach is most compelling when a compact chemically active region is embedded in a large polarizable environment. EFP's distributed multipoles, self-consistent induction, and overlap-dependent short-range terms offer a physically richer environment than fixed-charge force fields, while retaining a classical cost that is independent of quantum-register size.

Practical deployment on hardware will require ansatz and measurement reduction. UCCSD circuits can be deep, and the number of Pauli terms grows rapidly with active-space size. Adaptive ansatz construction, orbital optimization, measurement grouping, and error mitigation are natural extensions \cite{grimsley2019,temme2017,li2017,verteletskyi2020}. Analytical gradients, larger-basis active spaces, excited-state solvers, and molecular-dynamics sampling are also needed before the method can address quantitative free energies or spectroscopy.

\section{Conclusions}
Q-EFP combines a quantum treatment of a chemically active region with an EFP description of the surrounding environment. Electrostatic and polarization potentials modify the active-region Hamiltonian, while the remaining EFP contributions are assembled classically. In three STO-3G statevector benchmarks, Q-EFP differs from matched CCSD/EFP calculations by $0.01$--$0.38$ kcal mol$^{-1}$ and reduces estimated register sizes by approximately 10--35-fold relative to explicit treatment of all atoms. These results validate the current embedding and software workflow. Combined with Q-EFMO real-space fragmentation and Q-FVO virtual-space reduction, Q-EFP can form the outer embedding layer of a hierarchical workflow for larger solvated clusters. Larger active spaces, broader chemical environments, and quantum-hardware tests remain necessary.

\section{Acknowledgments}
This work was supported by the Laboratory Directed Research and Development (LDRD) program at Ames National Laboratory and by the NSF National Quantum Virtual Laboratory (NQVL) program. V.-A. G. acknowledges support from the Laboratory Directed Research and Development Program of Oak Ridge National Laboratory (LOIS ID 12735), managed by UT-Battelle, LLC, for the U.S. Department of Energy. This manuscript has been authored by UT-Battelle, LLC, under Contract No. DE-AC05-00OR22725 with the U.S. Department of Energy. This work was also supported by the U.S. Department of Energy, Office of Science, through Ames National Laboratory under Contract No. DE-AC02-07CH11358. The authors acknowledge partial support by ORNL LDRD and VSO programs. The U.S. Government retains, and the publisher by accepting the article for publication acknowledges, a nonexclusive, paid-up, irrevocable, worldwide license to publish or reproduce the published form of this manuscript, or allow others to do so, for U.S. Government purposes. This research used resources of the Oak Ridge Leadership Computing Facility (Frontier; Director's Discretionary allocation CHM238) and the National Energy Research Scientific Computing Center under award m4621 (ERCAP0036406). Computational resources were also provided by Iowa State University. The authors thank the IBM Quantum Researchers Program for access to quantum-computing resources.

\section{Data and code availability}
Data supporting the reported calculations are available from the corresponding author upon reasonable request. GAMESS is available at \url{https://www.msg.chem.iastate.edu/gamess/}. Q-GAMESS is available at \url{https://github.com/fzahari/Q-GAMESS}.

\FloatBarrier

\end{document}